# Least Square Method Robustness of Computations
## What is not usually considered and taught

Vaclav Skala

Department of Computer Science and Engineering
Faculty of Applied Sciences, University of West Bohemia
CZ 306 14 Plzen, Czech Republic
http://www.VaclavSkala.eu

*Abstract* — There are many practical applications based on the Least Square Error (LSE) approximation. It is based on a square error minimization "on a vertical" axis. The LSE method is simple and easy also for analytical purposes. However, if data span is large over several magnitudes or non-linear LSE is used, severe numerical instability can be expected.

The presented contribution describes a simple method for large span of data LSE computation. It is especially convenient if large span of data are to be processed, when the "standard" pseudoinverse matrix is ill conditioned. It is actually based on a LSE solution using orthogonal basis vectors instead of orthonormal basis vectors. The presented approach has been used for a linear regression as well as for approximation using radial basis functions.

I. INTRODUCTION

Wide range of applications is based on approximation of acquired data and the LSE minimization is used, known also as a linear or polynomial regression. The regression methods have been heavily explored in signal processing and geometrical problems or with statistically oriented problems. They are used across many engineering fields dealing with acquired data processing. Several studies have been published and they can be classified as follows:

- "standard" Least Square Error (LSE) methods fitting data to a function $y = f(x)$, where $x$ is an independent variable and $y$ is a measured or given value,
- "orthogonal" Total Least Square Error (TLSE) fitting data to a function $F(x) = 0$, i.e. fitting data to some $d-1$-dimensional entity in this $d$-dimensional space, e.g. a line in the $E^2$ space or a plane in the $E^3$ space [1][6][8][21][22],
- "orthogonally Mapping" Total Least Square Error (MTLSE) methods for fitting data to a given entity in a subspace of the given space. However, this problem is much more complicated. As an example, we can consider data given in and we need to find an optimal line in $E^d$, i.e. one dimensional entity, in this $d$-dimensional space fitting optimally the given data. Typical problem: Find a line in the $E^d$ space that has the minimum orthogonal distance from the given points in this space. This algorithm is quite complex and solution can be found in [18].

It should be noted, that all methods above do have one significant drawback as values are taken in a squared value. This results to an artifact that small values do not have relevant influence to the final entity as the high values. Some methods are trying to overcome this by setting weights to each measured data [3]. It should be noted that the TLSE was originally derived by Pearson [16](1901). Deep comprehensive analysis can be found in [8][13][21][22]. Differences between the LSE a TLSE methods approaches are significant, see Fig.1.

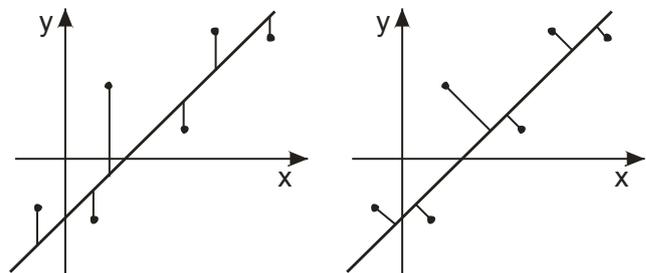

Fig. 1.a: Least Square Error    Fig.1.b: Total Least Square Error

In the vast majority the Least Square Error (LSE) methods measuring vertical distances are used. This approach is acceptable in the case of explicit functional dependences $f(x, y) = h$, resp. $f(x, y, z) = h$. However, it should be noted that a user should keep in a mind, that smaller differences than 1.0, will have significantly smaller weight than higher differences than 1.0 as the differences are taken in a square resulting to dependences in scaling of data approximated, i.e. the result will depend on physical units used, etc. The main advantage of the LSE method is that it is simple for fitting polynomial curves and it is easy to implement. The standard LSE method leads to over determined system of linear equations. This approach is also known as polynomial regression.

Let us consider a data set $\Omega = \{\langle x_i, y_i, f_i \rangle\}_{i=1}^n$, i.e. data set containing for $x_i, y_i$ and measured functional value $f_i$, and we want to find parameters $\boldsymbol{a} = [a, b, c, d]^T$ for optimal fitting function, as an example:

$$f(x, y, \boldsymbol{a}) = a + bx + cy + dxy \qquad (1)$$

Minimizing the vertical squared distance $D$, i.e.:

Research was supported by the and National Science Foundations (GACR) project No. 17-05534S.

$$D = \min_{a,b,c,d} \sum_{i=1}^{n} (f_i - f(x_i, y_i, \boldsymbol{a}))^2 =$$
$$\min_{a,b,c,d} \sum_{i=1}^{n} (f_i - (a + bx_i + cy_i + dx_iy_i))^2 \quad (2)$$

Conditions for an extreme are given as:
$$\frac{\partial f(x, y, \boldsymbol{a})}{\partial \boldsymbol{a}} = [1, x, y, xy]^T \quad (3)$$

Applying this on the expression of $D$ we obtain
$$\frac{\partial D}{\partial \boldsymbol{a}} \sum_{i=1}^{n} (f_i - (a + bx_i + cy_i + dx_iy_i)) \frac{\partial f(x, y, \boldsymbol{a})}{\partial \boldsymbol{a}} = 0 \quad (4)$$

It leads to conditions for $\boldsymbol{a} = (a, b, c, d)$ parameteters in the form of a linear system of equations $\boldsymbol{Ax} = \boldsymbol{b}$:

$$\boldsymbol{A} = \begin{bmatrix} n & \sum_{i=1}^{n} x_i & \sum_{i=1}^{n} y_i & \sum_{i=1}^{n} x_iy_i \\ \sum_{i=1}^{n} x_i & \sum_{i=1}^{n} x_i^2 & \sum_{i=1}^{n} x_iy_i & \sum_{i=1}^{n} x_i^2 y_i \\ \sum_{i=1}^{n} y_i & \sum_{i=1}^{n} x_iy_i & \sum_{i=1}^{n} y_i^2 & \sum_{i=1}^{n} x_iy_i^2 \\ \sum_{i=1}^{n} x_iy_i & \sum_{i=1}^{n} x_i^2 y_i & \sum_{i=1}^{n} x_iy_i^2 & \sum_{i=1}^{n} x_i^2 y_i^2 \end{bmatrix} \quad (5)$$

$$\boldsymbol{x} = [a, b, c, d]^T$$

$$\boldsymbol{b} = \left[ \sum_{i=1}^{n} f_i, \sum_{i=1}^{n} f_i x_i, \sum_{i=1}^{n} f_i y_i, \sum_{i=1}^{n} f_i x_i y_i \right]^T$$

The selection of bilinear form was used to show the LSE method application to a non-linear case, if the case of a linear function, i.e. $f(x, y, \boldsymbol{a}) = a + bx + cy$, the 4th row and column are to be removed. Note that the matrix $\boldsymbol{A}$ is symmetric and the function $f(\boldsymbol{x})$ might be more complex, in general.

Several methods for LSE have been derived [4][5][10], however those methods are sensitive to the vector $\boldsymbol{a}$ orientation and not robust in general as a value of $\sum_{i=1}^{n} x_i^2 y_i^2$ might be too high in comparison with the value $n$, which has an influence to robustness of a numerical solution. In addition, the LSE methods are sensitive to a rotation as they measure vertical distances. It should be noted, that rotational and translation invariances are fundamental requirements especially in geometrically oriented applications.

The LSE method is usually used for a small size of data and span of a domain is relatively small. However, in some applications the domain span can easily be over several decades, e.g. in the case of Radial Basis Functions (RBF) approximation for GIS applications etc. In this case, the overdetermined system can be difficult to solve.

## II. NUMERICAL STABILITY

Let us explore a simple example, when many points $\boldsymbol{x}_i \in E^2$, i.e. $\boldsymbol{x}_i = (x_i, y_i)$, are given with relevant associated values $b_i, i = 1, ..., n$. Expected functional dependency can be expressed (for a simplicity) as $y = a_1 + a_2 x + a_3 y$. The LSE leads to an overdetermined system of equations

$$\boldsymbol{A}^T \boldsymbol{A} \, \boldsymbol{\xi} = \boldsymbol{A}^T \boldsymbol{b} \quad (6)$$

where $\boldsymbol{b} = (b_1, ..., b_n)$, $\boldsymbol{\xi} = (\xi_1, ..., \xi_m)$ and $m$ is a number of parameters, $m < n$.

If the values $x_i, y_i$ over a large span, e.g. $x_i, y_i \in \langle 10^0, 10^5 \rangle$, the matrix $\boldsymbol{A}^T \boldsymbol{A}$ is extremely ill conditioned. This means that the reliability of a solution depends on the distribution of points in the domain. Situation gets worst when a non-linear polynomial regression is to be used and dimensionality of the domain is higher.

As an example, let us consider a simple case, when points form regular orthogonal mesh and values are generated using $R5$ distribution scheme (equidistant in a logarithmic scale) as $(x_i, y_i) \in \langle 10, 10^5 \rangle \times \langle 10, 10^5 \rangle$. It can be easily found using MATLAB that conditional number $cond(\boldsymbol{A}^T \boldsymbol{A}) \cong 10^{11}$.

In the following, we will show how the condition number might be decreased significantly using orthogonal basis vectors instead of the orthonormal ones.

## III. PROJECTIVE NOTATION AND GEOMETRY ALGEBRA

The LSE approximation is based on a solution of a linear system of equations, i.e. $\boldsymbol{Ax} = \boldsymbol{b}$. Usually the Euclidean representation is used. However if the projective space representation is used [19], it is transformed into homogeneous linear system of equations, i.e. $\boldsymbol{B\zeta} = \boldsymbol{0}$. Rewriting the Eq.(6), we obtain

$$\boldsymbol{B\zeta} = \boldsymbol{0} \quad (7)$$

where
$$\boldsymbol{B} = [-\boldsymbol{A}^T \boldsymbol{b} | \boldsymbol{A}^T \boldsymbol{A}]$$
$$\boldsymbol{\zeta} = (\zeta_0 : \zeta_1, ..., \zeta_m) \quad (8)$$

and $\xi_i = \zeta_i / \zeta_0$, $i = 1, ..., m$; $\zeta_0$ is the homogeneous coordinate in the projective representation, matrix $\boldsymbol{B}$ size is $m \times (m + 1)$. Now, a system of homogeneous linear equations is to me solved.

It can be shown that a system of homogeneous linear equations $\boldsymbol{Ax} = \boldsymbol{0}$ is equivalent to the extended cross-product, actually outer-product [19][20]. In general, solutions of the both cases $\boldsymbol{Ax} = \boldsymbol{0}$ and $\boldsymbol{Ax} = \boldsymbol{b}$, i.e. homogeneous and non-homogeneous system of linear equations, is the same and no division operation is needed as the extended cross-product (outer product) does not require any division operation at all. Applying this we get:

$$\boldsymbol{\zeta} = (\zeta_0 : \zeta_1, ..., \zeta_m) = \boldsymbol{\beta}_1 \wedge \boldsymbol{\beta}_2 \wedge ... \wedge \boldsymbol{\beta}_{m-1} \wedge \boldsymbol{\beta}_m \quad (9)$$

where
$$\boldsymbol{\beta}_i = [-b_{i0} : b_{i1}, ..., b_{im}]^T \quad i = 1, ..., m \quad (10)$$

The extended cross-product can be rewritten using determinant of $(m + 1) \times (m + 1)$ as

$$\boldsymbol{\zeta} = \det \begin{bmatrix} \boldsymbol{e}_0 & \boldsymbol{e}_1 & \boldsymbol{e}_2 & \cdots & \boldsymbol{e}_m \\ -b_{10} & b_{11} & b_{12} & \cdots & b_{1m} \\ \vdots & \vdots & \vdots & \ddots & \vdots \\ -b_{m0} & b_{m1} & b_{m2} & \cdots & b_{mm} \end{bmatrix} \quad (11)$$

where $\boldsymbol{e}_0$ are orthonormal basis vectors in the $m$-dimensional space. As a determinant is a multilinear, we can multiply any $j$ column by a value $q_j \neq 0$

$$\zeta' = \det \begin{bmatrix} e'_0 & e'_1 & e'_2 & \cdots & e'_m \\ -b'_{10} & b'_{11} & b'_{12} & \cdots & b'_{1m} \\ \vdots & \vdots & \vdots & \ddots & \vdots \\ -b'_{m0} & -b'_{m1} & -b'_{m2} & \cdots & -b'_{mm} \end{bmatrix} \quad (12)$$

where

$$e'_j = \frac{e_j}{q_j} \qquad b'_{*j} = \frac{b_{*j}}{q_j} \quad (13)$$

where $e'_j$ are orthogonal basis vectors in the $m$-dimensional space.

From the geometrical point of view, it is actually a "temporary" scaling on each axis including the units. Of course, a question remains – how to select the $q_j$ value. The $q_j$ is to be selected as

$$q_j = \max_{i=1,\dots,m}\{|b_{ij}|\} \quad (14)$$

where $j = 1,\dots,m$. Note that the matrix $\boldsymbol{B}$ is indexed as $(0,\dots,m) \times (0,\dots,m)$.

Applying this approach, we get a modified system

$$\zeta' = (\zeta'_0 : \zeta'_1, \dots, \zeta'_m) = \boldsymbol{\beta}'_1 \wedge \boldsymbol{\beta}'_2 \wedge \dots \wedge \boldsymbol{\beta}'_{m-1} \wedge \boldsymbol{\beta}'_m \quad (15)$$

where

$$\boldsymbol{\beta}'_i = [-b'_{i0} : b'_{i1}, \dots, b'_{im}]^T \quad (16)$$

where $\boldsymbol{\beta}'_i$ are coefficients of the matrix $\overline{\boldsymbol{B}}' = [-\boldsymbol{A}^T\boldsymbol{b}|\overline{\boldsymbol{A}^T\boldsymbol{A}}]$, i.e. modified matrix $\boldsymbol{B}$ as described above, for the orthogonal (not orthonormal) vector basis.

The approximated $f(x,y)$ value is computed as

$$f(x,y) = aq_1 + bq_2 x + cq_3 y \quad (17)$$

in the case of $f(x,y) = a + bx + cy$, or

$$f(x,y) = aq_1 + bq_2 x + cq_3 y + dq_4 xy \quad (18)$$

in the case $f(x,y) = a + bx + cy + dxy$ and similarly for the general case of a regression function $y = f(\boldsymbol{x},\boldsymbol{a})$.

The above presented modification is simple. However, what is the influence of this operation?

## IV. MATRIX CONDITIONALITY

Let us consider a recent simple example again, when points are generated from $(x_i, y_i) \in \langle 10, 10^5 \rangle \times \langle 10, 10^5 \rangle$. It can be found that conditional number $cond(\boldsymbol{A}^T\boldsymbol{A}) \cong 6.10^{10}$ using MATLAB, Fig.2, if $f(x,y) = a + bx + cy$ is used for the LSE.

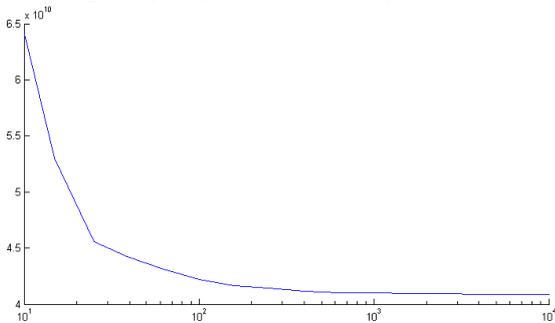

Fig.2: Conditionality histogram of the original matrix depending on number of data set size, i.e. number of points

Using the approach presented above, the conditional number was decreased significantly to $cond(\overline{\boldsymbol{A}^T\boldsymbol{A}}) \cong 2.10^6$.

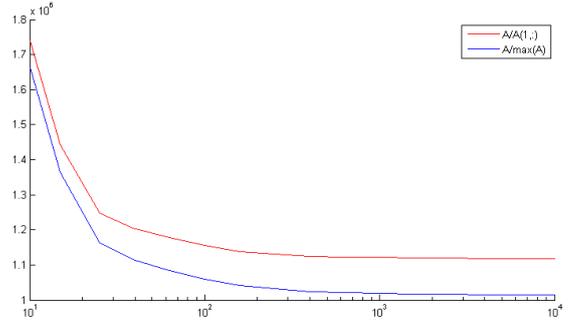

Fig.3: Conditionality of the modified matrix depending on number of data set size, i.e. number of points

Comparing the condition numbers of the original and modified matrices, we can see significant improvement of matrix conditionality as

$$v = \frac{cond(\boldsymbol{A}^T\boldsymbol{A})}{cond(\overline{\boldsymbol{A}^T\boldsymbol{A}})} \cong \frac{6.10^{10}}{2.10^6} = 3.10^4 \quad (19)$$

In the case of a little bit more complex function defined by Eq.(1), i.e. $f(x,y) = a + bx + cy + dxy$ we obtain

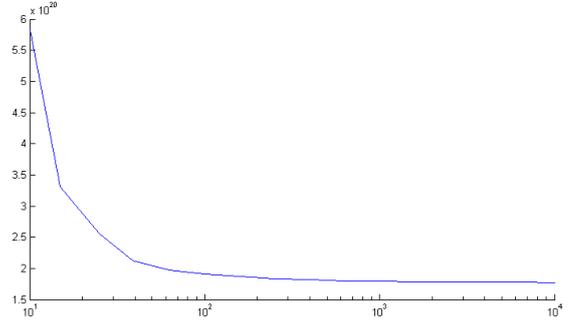

Fig.4: Conditionality of the original matrix depending on number of data set size, i.e. number of points

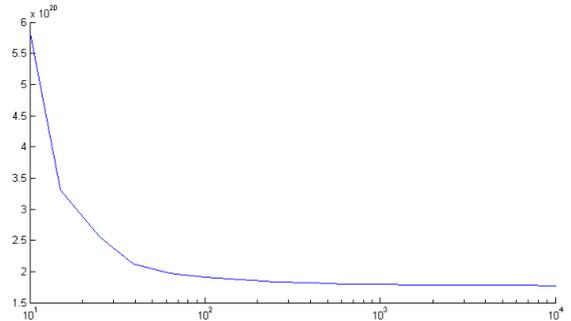

Fig.5: Conditionality of the modified matrix depending on number of data set size, i.e. number of points

In this case of the LSE defined by Eq.(1) the conditionality improvement is even higher, as

$$v = \frac{cond(\boldsymbol{A}^T\boldsymbol{A})}{cond(\overline{\boldsymbol{A}^T\boldsymbol{A}})} \cong \frac{6.10^{20}}{6.10^{11}} = 10^9 \quad (20)$$

It means that better numerical stability is obtained by a simple operation. All graphs clearly shows also dependency on a number of points used for the experiments (horizontal axis).

The geometric algebra brings also an interesting view on problems with numerical solutions. Let us consider vectors $\widehat{\boldsymbol{\beta}}_i$ with coordinates of points, i.e.

$$\widehat{\boldsymbol{\beta}}_i = [b_{i1}, \ldots, b_{im}]^T \quad i = 1, \ldots, m \quad (21)$$

Then $\widehat{\boldsymbol{\beta}}_i \wedge \widehat{\boldsymbol{\beta}}_j = \widehat{\boldsymbol{\gamma}}_{ij}$ defines a bivector, which is an oriented surface, given by two vectors in $m$-dimensional space and $\|\widehat{\boldsymbol{\gamma}}_{ij}\|$ gives the area represented by the bivector $\widehat{\boldsymbol{\gamma}}_{ij}$.

So, the proposed approach of introducing orthogonal basis functions instead of the orthonormal ones, enable us to "eliminate" influence of "small" bivectors in the original LSE computation and increase precision of numerical computation.

Of course, if the regression is to be applied, the influence of the $q_j$ values must be applied. By the presented approach we actually got values $\zeta_i'$ using the orthogonal basis vectors instead of orthonormal. It means, that the estimated value by a regression, using recent simple example, is

$$f(x, y) = q_1 a_1 + q_2 a_2 x + q_3 a_3 y \quad (22)$$

## V. Least Square Method with Polynomials

In the case of the least square approximation, we want to minimize using a polynomial of degree $n$.

$$\min_{P_n(x)} \|f(x) - P_n(x)\|$$
$$P_n(x) = \sum_{i=0}^{k} a_i x^i \quad (23)$$

The $L_2$ norme of a function $f(x)$ an an interval $\langle a, b \rangle$ is defined

$$\|f(x)\| = \sqrt{\left(\int_a^b f(x) dx\right)^2} \quad (24)$$

Minimizing square of the distance of a function of $k + 1$ parameters $\varphi(\boldsymbol{a}) = \varphi(a_0, \ldots, a_n)$ and using "per-partes" rule, we obtain

$$\varphi(\boldsymbol{a}) = \int_a^b [f(x) - P_n(x)]^2 dx$$
$$= \int_a^b [f(x)]^2 dx - 2 \sum_{i=0}^{n} a_i \int_a^b x_i f(x) dx \quad (25)$$
$$+ \sum_{i=0}^{n} \sum_{j=0}^{n} a_i a_j \int_a^b x^{i+j} dx$$

For a minimum a vector condition

$$\frac{\partial \varphi(\boldsymbol{a})}{\partial \boldsymbol{a}} = \boldsymbol{0} \quad (26)$$

must be valid. It leads to conditions

$$\frac{\partial \varphi(\boldsymbol{a})}{\partial a_k} = 0 - 2 \int_a^b x^k f(x) dx + \sum_{i=0}^{n} a_i \int_a^b x^{i+k} dx$$
$$+ \sum_{i=0}^{n} a_j \int_a^b x^{j+k} dx \quad (27)$$

and by simple algebraic manipulations we obtain:

$$2 \left[ -\int_a^b x^k f(x) dx + \sum_{i=0}^{n} a_i \int_a^b x^{i+k} dx \right] = 0 \quad (28)$$

and therefore

$$\sum_{i=0}^{n} a_i \int_a^b x^{i+k} dx = \int_a^b x^k f(x) dx \quad (29)$$

where $k = 1, \ldots, n$.

It means that the LSE problem is the polynomial (what has been expected)

$$P_n(x) = \sum_{i=0}^{k} a_i x^i \quad (30)$$

However, there is a direct connection with well known Hilbert's matrix. It can be shown that elements of the Hilbert's matrix $(H_{n+1}(a, b))_{i,k}$ of the size $(n + 1) \times (n + 1)$ are equivalent to

$$(H_{n+1}(a, b))_{i,k} = \int_a^b x^{i+k} dx = \frac{1}{1 + i + k} \quad (31)$$

If interval $\langle a, b \rangle = \langle 0, 1 \rangle$ is used, standard Hilbert's matrix $\boldsymbol{H}_n(0,1)$ is obtained, which is extremely ill-conditioned.

## VI. Hilbert's Matrix Conditionality

We should answer a question, how the conditional number of the Hilbert's matrix can be improved if orthogonal basis is used instead of orthonormal one as an experimental test.

A simple experiment can prove that the proposed method does not practically change the conditionality of the Hilbert's matrix $\boldsymbol{H}_n(0,1)$. However, as the LSE approximation is to be used for large span of data, it is reasonable to consider a general case and explore conditionality of the $\boldsymbol{H}_n(a, b)$ matrix, e.g. $\boldsymbol{H}_5(0, b)$, for demonstration.

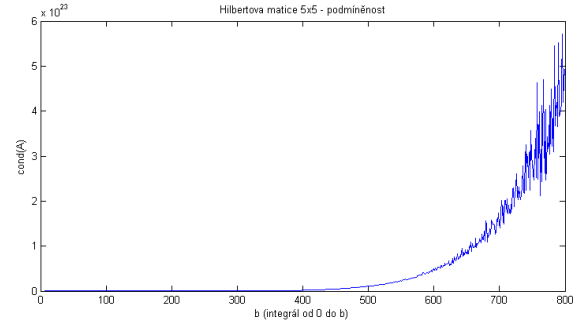

Fig.6: Conditionality of the $\boldsymbol{H}_5(0, b)$ for different values of $b$ using MATLAB (numerical problems can be seen for $b > 650$)

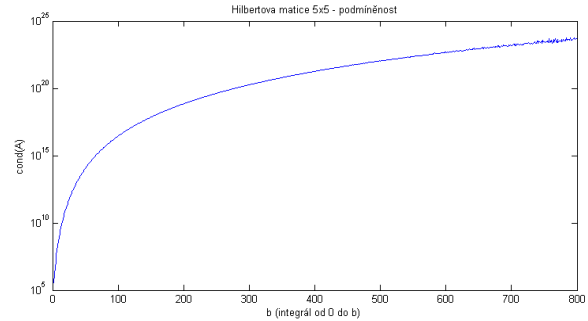

Fig.7: Conditionality of the $\boldsymbol{H}_5(0, b)$ for different values of $b$ using logarithmic scaling for vertical axis

It can be seen, that $cond(\boldsymbol{H}_5(0,800)) = 6.10^{23}$. If the proposed approach is applied $cond(\overline{\boldsymbol{H}_5(0,800)}) = 2,5.10^{14}$ for the modified matrix, Fig.8 - Fig.9.

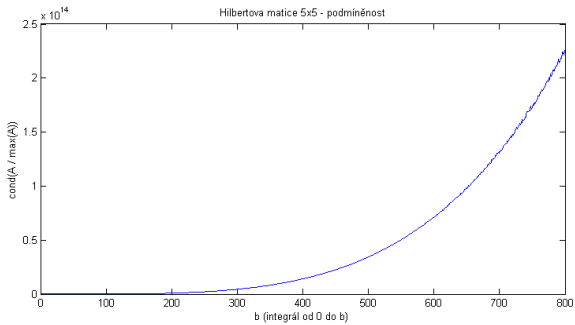

Fig.8: Conditionality of the modified $H_5(0,b)$

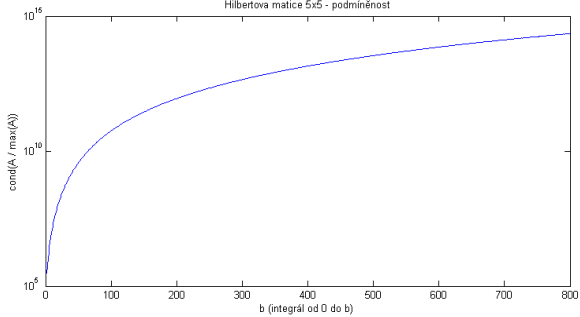

Fig.9: Conditionality of the modified $H_5(0,b)$ using logarithmic scaling for vertical axis

It means that the conditionality improvement

$$v = \frac{cond(H_5(0,800))}{cond(H_5(0,800))} \cong \frac{6.10^{23}}{2,5.10^{14}} \approx 10^9 \qquad (32)$$

This is a similar ratio as for the simple recent examples.

A change of the size of bivectors $\|\boldsymbol{\beta}_i \wedge \boldsymbol{\beta}_j\|$ can be used as a practical result using RBF approximation, which changes from the interval $\langle eps, 10^{10} \rangle$ to $\langle eps, 8.10^2 \rangle$, which significantly increases robustness of the RBF approximation, Fig.10.

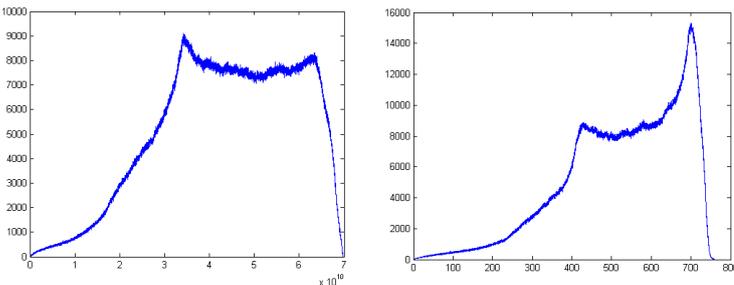

Fig.10: Bivector histogram sizes for original LSE matrix and modified one

The proposed approach has been used for St.Helen's volcano approximation by 10 000 points instead of 6 743 176 original points, see Fig.11.

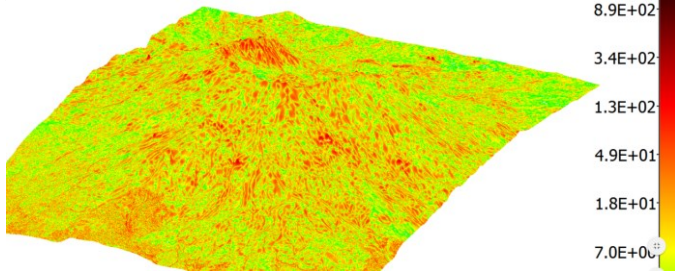

Fig.11: LSE approximation error with RBF approximation of St.Helen's (image generated in MATLAB by Michal Smolik)

## VII. CONCLUSIONS

The proposed method of application orthogonal vector basis instead of the orthonormal one decreases conditional number of a matrix used in the least square method. This approach increases robustness of a numerical solution especially when domain data range is high. It can be used also for solving systems of linear equations in general, e.g. if radial basis function interpolation or approximation is used.

### ACKNOWLEDGMENT

The author would like to thank to colleagues at the University of West Bohemia in Plzen for fruitful discussions and to anonymous reviewers for their comments and hints, which helped to improve the manuscript significantly. Special thanks belong to Zuzana Majdišová and Michal Šmolík for independent experiments, images and generation in MATLAB.